\documentclass[aps,prc,letterpaper,11pt,twoside,tightenlines,nofootinbib,showpacs,preprint]{revtex4}
\usepackage{graphicx}
\usepackage{epsfig,float}
\usepackage[sort&compress]{natbib}
\usepackage{subfigure}
\usepackage{amsmath}
\usepackage{amsfonts}
\usepackage{cancel}
\usepackage{lmodern,dsfont}
\usepackage{amssymb}
\usepackage{hyperref}
\usepackage{multirow}

\usepackage{color}

\begin{document}
\graphicspath{{./figure/}}

\title{\boldmath States of $\rho D^* \bar D^*$ with $J=3$ within the Fixed Center Approximation to the Faddeev equations }

\author{M. Bayar}
\affiliation{Department of Physics, Kocaeli University, 41380, Izmit, Turkey}
\affiliation{Departamento de F\'{\i}sica Te\'orica and IFIC, Centro Mixto Universidad de
Valencia-CSIC Institutos de Investigaci\'on de Paterna, Aptdo.
22085, 46071 Valencia, Spain}
\author{ Xiu-Lei Ren}
\affiliation{School of Physics and Nuclear Energy Engineering and International
Research Center for Nuclei and Particles in the Cosmos,
Beihang University, Beijing 100191, China}
\affiliation{Departamento de F\'{\i}sica Te\'orica and IFIC, Centro Mixto Universidad de
Valencia-CSIC Institutos de Investigaci\'on de Paterna, Aptdo.
22085, 46071 Valencia, Spain}
\author{E. Oset}
\affiliation{Departamento de F\'{\i}sica Te\'orica and IFIC, Centro Mixto Universidad de
Valencia-CSIC Institutos de Investigaci\'on de Paterna, Aptdo.
22085, 46071 Valencia, Spain}
\date{\today}

\begin{abstract}
 We study the interaction of the a $\rho$ and $D^*$, $\bar D^*$ with spins aligned using the Fixed Center Approximation to the Faddeev equations. We select a cluster of $D^* \bar D^*$, which is found to be bound in $I=0$ and can be associated to the X(3915), and let the $\rho$ meson orbit around the $D^*$ and $\bar D^*$. In this case we find an $I=1$ state with mass around 4340 MeV and narrow width of about 50 MeV. We also investigate the case with a cluster of $\rho D^*$ and let the $\bar D^*$ orbit around the system of the two states. The $\rho D^*$ cluster is also found to bind and leads to the $D^*_2(2460)$ state. The addition of the extra $\bar D^*$ produces further binding and we find, with admitted uncertainties, a state of $I=0$ around 4000 MeV, and a less bound narrow state with $I=1$ around 4200 MeV.  
\end{abstract}

\maketitle

\section{Introduction}

In hadron physics the existence of the three body resonances is drawing much attention for a long time. The  proper analysis of the three hadron system can be tackled by combining the Faddeev equations \cite{Faddeev:1960su} and chiral dynamics \cite{weinberg,gasser,ulfchiral}. However it is quite difficult to solve exactly the Faddeev equations. Recently, the combination of Faddeev equations and chiral dynamics was used to investigate for two meson-one baryon systems \cite{MartinezTorres:2007sr} and also for three mesons systems \cite{MartinezTorres:2008gy,MartinezTorres:2009xb}. On the other hand, the fixed center approximation (FCA) to the Faddeev equations is technically very simple and a powerful method to explore three hadron systems. This method is especially suitable to study the system in which two of the three particles are bound forming a cluster and this cluster is not much altered by the collision of the third particle \cite{MartinezTorres:2010ax,Bayar:2011qj,Bayar:2012rk}. This method has proved to be rather reliable for cases like $K$-deuteron scattering very close to threshold \cite{Toker:1981zh,Gal:2006cw}. In recent years the FCA to Faddeev equations has been successfully applied to the study of many three body interactions. For example in the work of the Ref. \cite{Xie:2011uw} the $\pi-(\Delta \rho)_{N_{(5/2)^{-}}(1675)}$  was analyzed by means of the FCA to the Faddeev equations in which the authors give a reasonable explanation for the $\Delta_{(5/2)^{+}}(2000)$ puzzle. Likewise, the $N\bar{K}K$ system was investigated using the Faddeev equations under the FCA in Ref. \cite{Xie:2010ig} and the results are in good agreement with the variational estimation \cite{Jido:2008kp} and also the full Faddeev calculation \cite{MartinezTorres:2008kh,MartinezTorres:2010zv}.

In the present paper we want to study the $  \rho D^* \bar{D}^* $ system. The reason is that the vector-vector interaction is found to be very strong, particularly in the $J=2$ channel \cite{Molina:2008jw, Geng:2008gx}. Due to this, it was possible to see that multi-$\rho$ states with the spins parallel were  bound, although the width was increasing with the number of  $\rho$'s \cite{Roca:2010tf}. The masses and widths obtained were in good agreement with experimental data. The same occurred with $K^{*}$ multi-$\rho$ states in \cite{YamagataSekihara:2010qk} and with $D^{*}$ multi-$\rho$ states in \cite{Xiao:2012dw}. In the case of  $K^{*}$ multi-$\rho$ states one finds good agreement with experiment, but in the case of $D^{*}$ multi-$\rho$ states only predictions were made and experimental counterparts have not yet been observed.

The $  \rho D^* \bar{D}^* $ system is new and has not been studied so far. Yet, studies done for $D^* \bar{D}^* $ interaction in \cite{DsDBs} and $   D^* \rho$ interaction in \cite{rhoDstr} have already set the grounds to tackle this interesting system with hidden charm, and we wish to study it here.

\section{\label{formalism} FORMALISM}

In this section we describe the calculation of the three body interaction of the  $  \rho D^* \bar{D}^* $ system in s-wave and all the spins aligned.  In the work of \cite{rhoDstr}, the $\rho D^* $ interaction was studied using the hidden gauge formalism \cite{Bando:1987br,Bando:1984ej,Meissner:1987ge}. In \cite{rhoDstr} the authors found strong attraction in $I=1/2$, $J=2$ which  corresponds to the tensor state $D_{2}^{*}(2460)$ with $I(J^{P})=\frac{1}{2}(2^{+})$. Similarly, it was also found in Ref. \cite{DsDBs}  that the resonance $X(3915)$ ($I^{G}(J^{PC})=0^{+}(2^{++})$) could be understood as a molecule made of  $ D^*$  and  $ \bar{D}^*$ mesons  with strong attraction in the $I=0$, $J=2$ sector. 

We use the FCA to the Faddeev equations to study the $\rho D^* \bar{D}^*  $ system. This method is particularly well suited for the system in which a pair of particles clusters together and the cluster is not much modified by the third particle, like in the case of $D_{2}^{*}(2460)$ as a $\rho D^* $ cluster and $X(3915)$ as a $ D^* \bar{D}^*$ cluster.

For the technical details we proceed similarly to Ref. \cite{Bayar:2011qj} to express the FCA to the Faddeev equations. Hereinafter we are going to interpret the interaction of a particle $a_3$ with a cluster made of two particles, $a_1$ and $a_2$. In this work the particle $a_3$ will represent the $\rho$($\bar{D}^*$) which scatters from the cluster, $X(3915)$ ($D_{2}^{*}(2460)$) and $a_1$ and  $a_2$ are the $D^*$($\rho$) and $\bar{D}^*$($D^*$) which build up the cluster. 

\begin{center}
  \begin{figure}
  \resizebox{0.9\textwidth}{!}{%
  \includegraphics{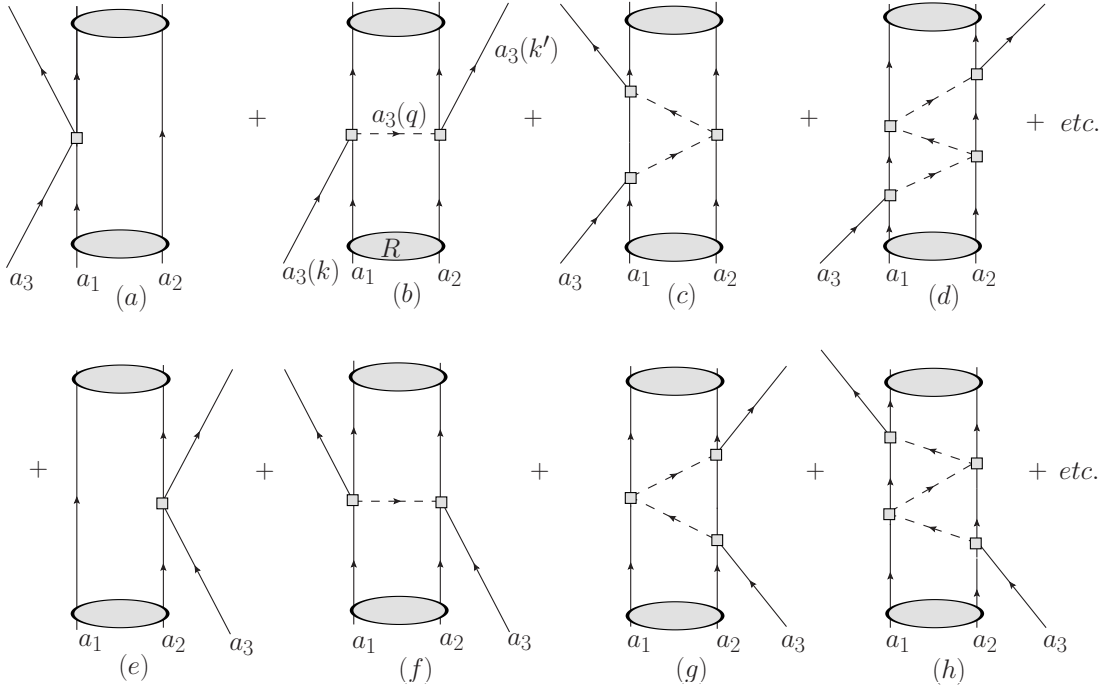}
  }
  \caption{ \label{fig:FaddEq}
  Diagrammatic representation of the fixed center approximation to Faddeev equations.}
  \end{figure}
  \end{center}

The diagramatic representation of the FCA to the Faddeev equations is shown in Fig. \ref{fig:FaddEq}. The particle $a_3$ rescatters repeatedly with the components of the cluster. In Fig. \ref{fig:FaddEq} the thick squared dots represent the unitary scattering amplitudes with coupled channel for the interaction of  particle  $a_3$ with particle $a_1$ ($t_1$) and  $a_2$ ($t_2$), respectively, which will be discussed later. In order to write the equations for the total three body scattering amplitude, we define two partition functions $T_1$, $T_2$ which sum all diagrams of the series of Fig. \ref{fig:FaddEq} which begin with the interaction of particle $a_3$ with particle $a_1$ of the cluster ($T_1$), or with particle $a_2$ ($T_2$). Then the FCA equations can be written as a system of coupled equations: 

   \begin{eqnarray}  
  T_1 &=&t_1 + t_1 G_0 T_2, \nonumber \\
  T_2&=&t_2 + t_2 G_0 T_1, \nonumber \\
  T &=&  T_1 + T_2  
 \label{eq:ScatAmp} 
  \end{eqnarray}
  where $G_0$ is the Green function for the propagator of particle $a_3$ between the particles  $a_1$ and  $a_2$ which is discussed later on. 
  
  The scattering amplitude $ \langle  \rho D^{*}\bar{D}^{*}  \vert t \vert  \rho D^{*}\bar{D}^{*}  \rangle $ for the single scattering contribution is obtained in terms of the two-body amplitudes $t_{1}$, $t_{2}$ derived in Refs. \cite{rhoDstr,DsDBs}. First we explicitly determine the case of $\rho (D^{*}\bar{D}^{*}) $ in which  $I_{D^{*}\bar{D}^{*}}=0$ and the total isospin of the three body system $I_{\rho (D^{*}\bar{D}^{*})}=1$. Using the nomenclature $\mid \rho, I, I_{z} \rangle \otimes  \mid D \bar{D}, I, I_{z} \rangle  $ we obtain

  \begin{eqnarray}
   \langle \rho (D^{*}\bar{D}^{*}) \vert t \vert \rho (D^{*}\bar{D}^{*}) \rangle  &=&\langle \rho^{+} (D^{*}\bar{D}^{*})^{I=0} \mid  (\hat{t}_{\rho D^{*}}+\hat{t}_{\rho \bar{D}^{*}} ) \mid \rho^{+} (D^{*}\bar{D}^{*})^{I=0}  \rangle \nonumber\\
  &=& - \langle +1 \mid  \otimes \frac{1}{\sqrt{2}} \Big( \langle \frac{1}{2}, \frac{1}{2} \mid   \langle \frac{1}{2}, -\frac{1}{2} \mid - \langle \frac{1}{2}, -\frac{1}{2} \mid
   \langle \frac{1}{2}, \frac{1}{2} \mid \Big) (\hat{t}_{\rho D^{*}}+\hat{t}_{\rho \bar{D}^{*}} )\nonumber\\ && 
   (-) \mid  +1\rangle \otimes \frac{1}{\sqrt{2}} \Big( \mid \frac{1}{2}, \frac{1}{2} \rangle  \mid \frac{1}{2}, -\frac{1}{2} \rangle - \mid \frac{1}{2}, -\frac{1}{2} \rangle
   \mid \frac{1}{2}, \frac{1}{2} \rangle \Big)  \nonumber\\
  &=& \Big( \frac{2}{3} t_{\rho D^{*}}^{I=3/2}+ \frac{1}{3} t_{\rho D^{*}}^{I=1/2}\Big)+\Big( \frac{2}{3} t_{\rho \bar{D}^{*}}^{I=3/2}+ \frac{1}{3} t_{\rho \bar{D}^{*}}^{I=1/2}\Big)
  \label{IsospinrhoDsDbS}
   \end{eqnarray}
where 
\begin{eqnarray}
  \mid D^{*}\bar{D}^{*} \rangle^{I=0} =\frac{1}{\sqrt{2}} \mid \frac{1}{2}, -\frac{1}{2} \rangle-\frac{1}{\sqrt{2}} \mid -\frac{1}{2}, \frac{1}{2} \rangle
  \end{eqnarray}
  with the nomenclature $\mid I_{z}\rangle$ for the $\rho$ meson and $\mid I_{z_{1}}, I_{z_{2}} \rangle$ for the $ D^{*} \bar{D}^{*}$ system. 
  
  Second we write the case of $ \bar{D}^{*} (\rho D^{*})  $ where  $I_{\rho D^{*}}=1/2$ and the total isospin of the three body system $I_{  \bar{D}^{*} (\rho D^{*}) }=0$ or $I_{  \bar{D}^{*} (\rho D^{*})}=1$.
  
   For the total isospin $I=1 $ case 

 \begin{eqnarray}
  \langle  \bar{D}^{*} (\rho D^{*}) \vert t \vert \bar{D}^{*}(\rho D^{*}) \rangle &=&\langle \bar{D}^{*}  (\rho D^{*})^{I=1/2} \mid  (\hat{t}_{D^{*} \rho }+\hat{t}_{D^{*} \bar{D}^{*}} ) \mid \bar{D}^{*} (\rho D^{*})^{I=1/2} \rangle  \nonumber\\
  &=&\Big[ \dfrac{1}{\sqrt{2}} \langle \frac{1}{2} \mid  \otimes \Big( \frac{1}{\sqrt{3}}  \langle 0, -\frac{1}{2} \mid  -\sqrt{\frac{2}{3}} \langle -1, \frac{1}{2} \mid \Big ) \nonumber\\ && + \frac{1}{\sqrt{2}} \langle -\frac{1}{2} \mid \otimes 
   \Big ( \sqrt{\dfrac{2}{3}} \langle 1,-\frac{1}{2} \mid -\dfrac{1}{\sqrt{3}} \langle 0, +\frac{1}{2} \mid \Big)\Big] \nonumber\\ && \Big(\hat{t}_{D^{*} \rho}+\hat{t}_{D^{*} \bar{D}^{*}} \Big)\nonumber\\ &&
  \Big[\dfrac{1}{\sqrt{2}} \mid \frac{1}{2} \rangle  \otimes \Big( \frac{1}{\sqrt{3}}  \mid 0, -\frac{1}{2} \rangle -\sqrt{\frac{2}{3}} \mid -1, \frac{1}{2} \rangle \Big ) \nonumber\\ &&+ \frac{1}{\sqrt{2}} \mid -\frac{1}{2} \rangle \otimes 
   \Big ( \sqrt{\dfrac{2}{3}} \mid 1, -\frac{1}{2} \rangle -\dfrac{1}{\sqrt{3}} \langle 0, +\frac{1}{2} \rangle \Big)\Big]\nonumber\\
  &=& \Big( \frac{8}{9} t_{D^{*} \rho }^{I=3/2}+ \frac{1}{9} t_{D^{*} \rho}^{I=1/2}\Big)+\Big( \frac{2}{3} t_{D^{*} \bar{D}^{*}}^{I=1}+ \frac{1}{3} t_{D^{*} \bar{D}^{*}}^{I=0}\Big)
  \label{eq:Spin1}
   \end{eqnarray}
with the nomenclature    $ \mid \bar{D}^{*}, I_{z} \rangle \otimes  \mid \rho D^{*}, I_{z_{1}}, I_{z_{2}} \rangle $. In the case of the total  isospin $I=0$ case we similarly derive  

\begin{eqnarray}
  \langle  \bar{D}^{*} (\rho D^{*}) \vert t \vert \bar{D}^{*}(\rho D^{*}) \rangle &=& \Big( t_{D^{*} \rho }^{I=1/2} \Big)+ \Big( t_{D^{*} \bar{D}^{*}}^{I=1}\Big).\label{eq:Spin2}
   \end{eqnarray}

 Since we use the normalization of Mandl and Shaw \cite{mandl}, which has different weight factors for the particle fields, we need to consider how these factors are adapted to the present problem. It is easy to do this compairing the single scattering, double scattering and full scattering amplitudes. In this case, following the field normalization of Ref. \cite{mandl}, we can obtain the S matrix for the single scattering diagram (Fig. \ref{fig:FaddEq} (a) and (e)),
  
   \begin{eqnarray} S_{1}^{(1)}
  &=&-i t_{1} \frac{1}{{\cal V}^2} (2\pi)^4 \delta^4 (k + k_{R} - k^{'} - k_{R}^{'}) \nonumber \\
  &\times &  \frac{1}{\sqrt{2\omega_{a_3}}} \frac{1}{\sqrt{2\omega_{a_3}^{'}}}
  \frac{1}{\sqrt{2\omega_{a_1}}} \frac{1}{\sqrt{2\omega_{a_1}^{'}}},
  \label{eq:S1}
  \end{eqnarray}
  
   \begin{eqnarray} S_{2}^{(1)}
  &=&-i t_{2} \frac{1}{{\cal V}^2} (2\pi)^4 \delta^4 (k + k_{R} - k^{'} - k_{R}^{'}) \nonumber \\
  &\times &  \frac{1}{\sqrt{2\omega_{a_3}}} \frac{1}{\sqrt{2\omega_{a_3}^{'}}}
  \frac{1}{\sqrt{2\omega_{a_2}}} \frac{1}{\sqrt{2\omega_{a_2}^{'}}},
  \label{eq:S2}
  \end{eqnarray}
where the momentum $k(k')$, the on-shell energy $\omega (\omega^{'})$ refer to the initial (final) particles, respectively, and ${\cal V}$ is the volume of the box where the states are normalized to unity. In Eqs. (\ref{eq:S1}) and (\ref{eq:S2}), $t_{1}$, $t_{2}$ correspond to the first and second terms of the right hand side of Eqs. (\ref{eq:Spin1}) and (\ref{eq:Spin2}).

Likewise we have the $S$-matrix for the double scattering diagram as (Fig.~\ref{fig:FaddEq} $(b)$ or $(f)$) 

  \begin{eqnarray}
  S^{(2)}&=&
  -i  (2\pi)^4  \frac{1}{ {\cal V}^2 }  \delta^4 (k + k_{R} - k^{'} - k_{R}^{'}) \nonumber \\
  &\times& 
  \frac{1}{ \sqrt{2\omega_{a_3}} } \frac{1}{ \sqrt{2\omega_{a_3}^{'}} }
  \frac{1}{ \sqrt{2\omega_{a_1}} } \frac{1}{ \sqrt{2\omega_{a_1}^{'}} }
  \frac{1}{ \sqrt{2\omega_{a_2}} } \frac{1}{ \sqrt{2\omega_{a_2}^{'}} } \nonumber \\
  &\times&
  \int \frac{d^3 q}{(2\pi)^3} F_{R}(q) \frac{1}{ q^{0^{2}} -\vec{q}~^2 - m_{_{a_3}}^2 + i\epsilon }t_{1} t_{2} ,
  \label{eq:Stbody}
  \end{eqnarray}
where $F_{R}(q)$ is the form factor of the cluster which represents essentially the Fourier transform of its wave function. The derivation of the form factors can proceed similarly to Refs. 
\cite{Roca:2010tf,YamagataSekihara:2010pj}, where one can also read further discussions and interpretation. 
The form factor for s-wave functions is given by

 \begin{eqnarray}
  F_{R} (q) &=& \frac{1}{{\cal N}}
  \int_{ \stackrel{p < k_{{ max} }}{|{\vec{p} - \vec{q}}| < k_{{max}}}}
  d^3p ~ \frac{1}{2 \omega_{a_1}(\vec{p~})}   ~\frac{1}{2 \omega_{a_2}(\vec{p~})}
  ~ \frac{1}{M_{R}-\omega_{a_1}(\vec{p~})-\omega_{a_2}(\vec{p~})} \nonumber \\
  &\times &  \left( \frac{1}{2 \omega_{a_1}(\vec{p} -\vec{q~})}\right) \left( \frac{1}{2 \omega_{a_2}(\vec{p} -\vec{q~})}\right)
  \frac{1}{M_{R}-\omega_{a_1} ( \vec{p}-\vec{q~})-\omega_{a_2} ( \vec{p}-\vec{q~}) },
  \label{eq:formfactor}
  \end{eqnarray}
with the normalization ${\cal N}$
  \begin{eqnarray}
  {{\cal N }}
  =
  \int_{p < k_{{max}}} d^3p \left[ \frac{1}{2\omega_{a_1} (\vec{p~})}~\frac{1}{2\omega_{a_2}(\vec{p~})}~ \frac{1}{M_{R}-  \omega_{a_1}(\vec{p~})-\omega_{a_2}(\vec{p~})} \right]^2  \nonumber \\
  \label{eq:normformf}
  \end{eqnarray}
where $\omega_{a_1}$  and $\omega_{a_2}$ are the energies of the particles $a_1$, $a_2$, and $k_{{max}}$ is a cutoff that regularizes the integral of Eqs. (\ref{eq:formfactor}) and (\ref{eq:normformf}). This
cutoff is the same one needed in the regularization of the
loop function of the two particle propagators in the study of the interaction of the two particles of the cluster \cite{YamagataSekihara:2010pj}. In this work we take  the cutoff $k_{{max}}=1200$ MeV, the same one used to generate the $D_{2}^{*}(2460)$  \cite{rhoDstr}.

Similarly, the full three body $S$-matrix for scattering of particle $a_3$ with the cluster is given by  
  
   \begin{eqnarray}
  S  &=&
  -i T \frac{1}{{\cal V}^2} (2\pi)^4 \delta^4 (k + k_{R} - k^{'} - k_{R}^{'}) \nonumber \\
  && \frac{1}{ \sqrt{ 2\omega_{a_3} } } \frac{1}{ \sqrt{ 2\omega_{a_3}^{'} } }
  \frac{1}{ \sqrt{ 2\omega_{R}} } \frac{1}{ \sqrt{ 2\omega_{R}^{'}} } .
  \label{eq:ST}
  \end{eqnarray}
Compairing this equation with Eqs. (\ref{eq:S1})  and (\ref{eq:S2}), we introduce convenient factors in the elementary amplitudes: 

\begin{eqnarray}
 \tilde{t}_{1(2)}=\frac{2~M_{R}}{2~m_{a_{1(2)}}} t_{1(2)}.
 \end{eqnarray}
with $m_{a_1}$, $m_{a_2}$ and $M_{R}$ the masses of the particle $a_1$, $a_2$ and the cluster respectively, where we have taken the approximations, suitable for bound states,
 $ \frac{1}{\sqrt{2 \omega_{a_{1(2)}}}}=\frac{1}{\sqrt{2 m_{a_{1(2)}}}} $.

  Finally solving the set of equations for the  FCA to the Faddeev equations, Eqs. (\ref{eq:ScatAmp}), we obtain 
  
  \begin{equation}
T=T_1+T_2=\frac{ \tilde{t}_{1}+\tilde{t}_{2} + 2 \tilde{t}_{1} \tilde{t}_{2} G_0}{1-\tilde{t}_{1} \tilde{t}_{2} G_0^2}.
\label{Eq:totalT}
\end{equation}

Note that the argument of the total amplitude $T$ is regarded as a function of the total invariant mass of the three body system whereas the argument of $t_{1(2)}$  is the invariant masses of the two body systems. 
In order to obtain the arguments $s_{1(2)}$ of the two body amplitude  we share the binding energy among the three particles, proportionally to their masses. Therefore the energy of the particles $a_1$,  $a_2$ and $a_3$ become

\begin{equation}
E_{a_3}=m_{a_3} \frac{\sqrt{s}}{(M_{R}+m_{a_3})}
\end{equation}

\begin{equation}
E_{a_1}=\frac{\sqrt{s}}{(M_{R}+m_{a_3})}\dfrac{m_{a_1}~ M_{R} }{(m_{a_1}+m_{a_2})}
\end{equation}
\begin{equation}
E_{a_2}=\frac{\sqrt{s}}{(M_{R}+m_{a_3})}\dfrac{m_{a_2}~ M_{R} }{(m_{a_1}+m_{a_2})}
\end{equation}

Hence the total energy of the two body system is evaluated as follows 

\begin{equation}
s_{1(2)}=(p_{a_3}+p_{a_1(a_2)})^{2}=\Big( \frac{\sqrt{s}}{M_{R}+m_{a_3}}\Big)^{2} (m_{a_3}+\dfrac{m_{a_1(a_2)}~M_{R}}{m_{a_1}+m_{a_2}})^{2}-\vec{P}_{a_2(a_1)}^{2}
\end{equation}
where the approximate value of $\vec{P}_{a_2(a_1)}$ is given by

\begin{equation}
\frac{\vec{P}_{a_2(a_1)}^{2}}{2~m_{a_2(a_1)}} \simeq B_{a_2(a_1)} \equiv \frac{m_{a_2(a_1)}~ M_{R}}{(m_{a_1}+m_{a_2})} \frac{(M_{R}+m_{a_3}-\sqrt{s})}{(M_{R}+m_{a_3})} 
\end{equation}
with $ B_{a_2(a_1)}$ the binding energy of the particle $a_2(a_1)$.

\begin{center}
  \begin{figure}
  \resizebox{0.8\textwidth}{!}{
  \includegraphics{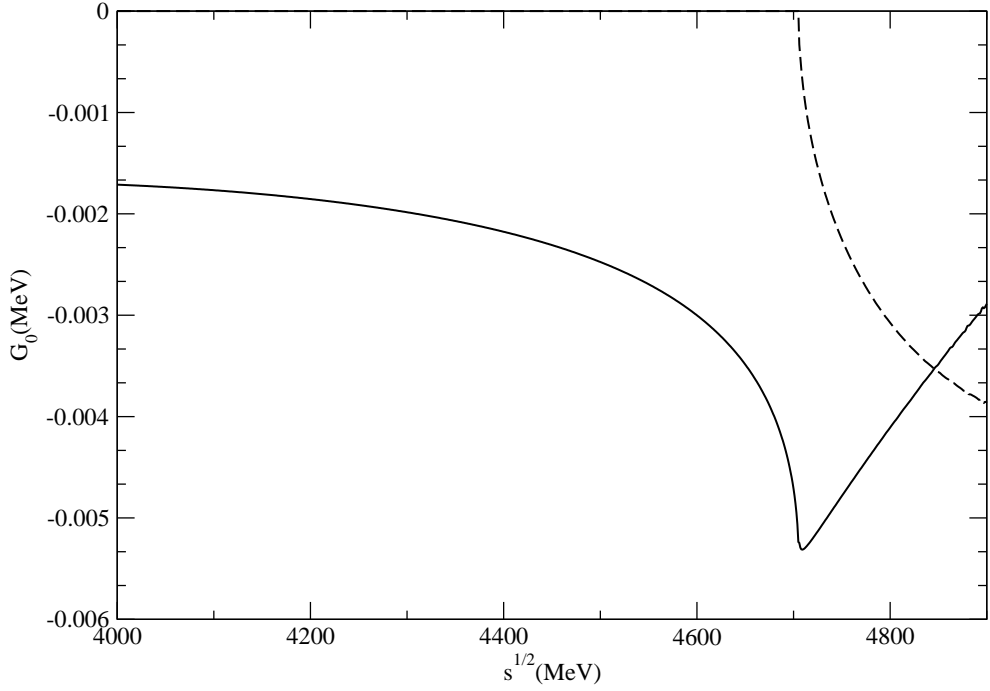}  }
  \caption{Real (solid line) and imaginary (dashed line) parts of $G_0$ function in $ \rho (D^{*}\bar{D}^{*})$}
  \label{fig:G0}
  \end{figure}
  \end{center}

As we stated before, the $G_{0}$ function is the propagator of the particle $a_3$ inside the cluster as follows

\begin{equation}
G_0=\frac{1}{2 M_{R}}\int\frac{d^3q}{(2\pi)^3}F_{R}(q)\frac{1}{{q^0}^2-\vec{q}\,^2-m_{a_3}^2+i\epsilon}.
\label{Eq:G0}
\end{equation}
where $M_{R}$ is the mass of the cluster, and $m_{a_3}$ the mass of the particle $a_3$. Here the energy of the propagator $q^{0}$ is determined at the three body rest frame as
\begin{equation}
q^0=\frac{s+m_{a_3}^2-M_{R}^2}{2\sqrt{s}}
\end{equation}
with $\sqrt{s}$ the rest energy of the three body system. As an example we depict in Fig. \ref{fig:G0} the real and imaginary parts of the $G_{0}$ function for the $  \rho (D^{*}\bar{D}^{*})$ system. The $G_{0}$ function has a similar shape for the $ \bar{D}^{*} (\rho D^{*}) $  system but with a different threshold. 

As we mentioned previously, in the evaluation of  the $  \rho D^{*} \bar{D}^{*} $ three body interaction, the use of the $ \rho D^{*}$, $\rho \bar{D}^{*}$ and $D^{*} \bar{D}^{*}$ unitarized amplitudes has crucial importance. These amplitudes were studied by the coupled channel Bethe-Salpeter equations in Refs. \cite{rhoDstr,DsDBs} and we use them here. In order to reproduce the $\rho D^{*} (\rho\bar{D}^{*})$ system  from the work of \cite{rhoDstr}, the coupled channels used are $\rho D^{*}$ and $\omega D^{*}$ for $I=\frac{1}{2}$ and $\rho D^{*}$  for $I=\frac{3}{2}$ case. In the case of  the $D^{*}\bar{D}^{*}$ system \cite{DsDBs, Geng:2008gx} there are $10$ coupled channels, $D^{*}\bar{D}^{*}$, $K^{*}\bar{K}^{*}$, $\rho \rho$, $\omega \omega$, $\phi \phi$, $J/ \Psi J/ \Psi$, $\omega J/ \Psi$, $\phi J/ \Psi$, $\omega \phi $ and $D_{s}^{*}\bar{D}_{s}^{*}$ for $I=0$ and six coupled channels $D^{*}\bar{D}^{*}$, $K^{*}\bar{K}^{*}$, $\rho \rho$, $\rho \omega $, $\rho J/ \Psi$ and $ \rho \phi$ for $I=0$. 

Following the ideas of the coupled channels chiral unitary approach, the $VV$-two body scattering amplitude can be obtained using the Bethe-Salpeter equations in its on-shell factorized form as below

\begin{eqnarray}
 t&=&({\hat 1} - V{\hat G})^{-1} V \label{betaSalseq}
\end{eqnarray}
where the $V$ is a matrix of the interaction potentials between the channels, which is calculated from the hidden gauge Lagrangian \cite{Bando:1984ej, Bando:1987br, Meissner:1987ge}. The potential $V$ is a $10 \times 10$ matrix in $I=0$ and $6\times6$ matrix in $I=1$ with the amplitudes obtained from the coupled channels for $D^{*}\bar{D}^{*}$ case in $J=2$ \cite{DsDBs, Geng:2008gx}. In addition, in the case of the $ \rho D^{*}$ the potential $V$ is a $2 \times 2$ matrix given by \cite{rhoDstr}.

In Eq. (\ref{betaSalseq}) $\widehat{G}$ is a diagonal matrix of the loop function of two mesons in the $i$ channel

\begin{equation}
 \widehat{G}_{i}(P)=i \int \frac{d^{4}q}{(2 \pi)^{4}} \frac{1}{q^{2}-m_{1}+i \epsilon}\frac{1}{(P-q)^{2}-m_{2}+i \epsilon} 
 \label{Gloop}
\end{equation}
where $P$ is the four dimensional momentum of the two vector mesons determined at the rest frame, $P=(\sqrt{s},0)$, and  $m_{1}$ and $m_{2}$ are the masses of the vector mesons in the $i$ channel. In order to remove the ultraviolet divergence of the loop function, we use the dimensional regularization scheme and we get

  \begin{eqnarray}
   \widehat{G}_i(\sqrt{s})&=&{1 \over 16\pi ^2}\biggr( \alpha _i+Log{m_1^2 \over \mu ^2}+{m_2^2-m_1^2+s\over 2s}
  Log{m_2^2 \over m_1^2}+\nonumber \\ 
 & &{q_{i}\over \sqrt{s}}\Big( Log{s-m_2^2+m_1^2+2q_{i}\sqrt{s} \over -s+m_2^2-m_1^2+
  2q_{i}\sqrt{s}}+Log{s+m_2^2-m_1^2+2q_{i}\sqrt{s} \over -s-m_2^2+m_1^2+  2q_{i}\sqrt{s}}\Big)\biggr)
  \label{loop}
\end{eqnarray}
where $q_{i}$ is the three momentum of the two vector mesons determined at the center of mass frame evaluated as follows

\begin{eqnarray}
q_{i}&=&{\sqrt{(s-(m_1+m_2)^2)(s-(m_1-m_2)^2)}\over 2\sqrt{s}}. \label{trimomentum}
\end{eqnarray}

In Eq. (\ref{loop}),  $\mu$ is a regularization scale and $\alpha_{i}$ is the subtraction constant. We take $\mu=1500$ MeV and $\alpha=-1.74$ for the $ \rho D^{*}$ to reproduce $D_{2}^{*}(2460)$ resonance \cite{rhoDstr} and $\mu=1000$ MeV and $\alpha=-2.07$ in the channel including heavy mesons, $\alpha=-1.65$ in the channel including light mesons for $D^{*}\bar{D}^{*}$ to obtain the X(3915) state \cite{DsDBs}. In Eq. (\ref{eq:formfactor}), the cutoff regularizes the integral of the form factor. In the present work, we choose the cutoff $k_{max}=1200$ MeV both for the $D_{2}^{*}(2460)$ and  $X(3915)$,   which produces similar results as the using the chosen subtraction constants in the dimensional regularization.

    \section{\label{sec:results} Results} 
    
In this section we present the results obtained for the scattering amplitude of the $\rho D^{*}\bar{D}^{*} $ system in spin-3. The two-body  $ \rho D^{*}$, $\rho \bar{D}^{*}$ and $D^{*}\bar{D}^{*}$ systems were investigated by the coupled channel Bethe-Salpeter equations in  Refs. \cite{rhoDstr,DsDBs}.  As we stated before, the resonance $D_{2}^{*}(2460)$ was generated as a $ \rho D^{*}$ quasibound state or molecule in the isospin $1/2$ and spin-2. It was also found that the resonance $X(3915)$ is dynamically generated  in $I=0$ and spin-2 from $D^{*}\bar{D}^{*}$ scattering. Therefore, there are two possible cases of three-body scattering of the $\rho D^{*}\bar{D}^{*} $ system. One is the $D_{2}^{*}(2460)-\bar{D}^{*}$ and the other one is the $X(3915)- \rho$. 

\begin{center}
  \begin{figure}
  \resizebox{0.8\textwidth}{!}{
  \includegraphics{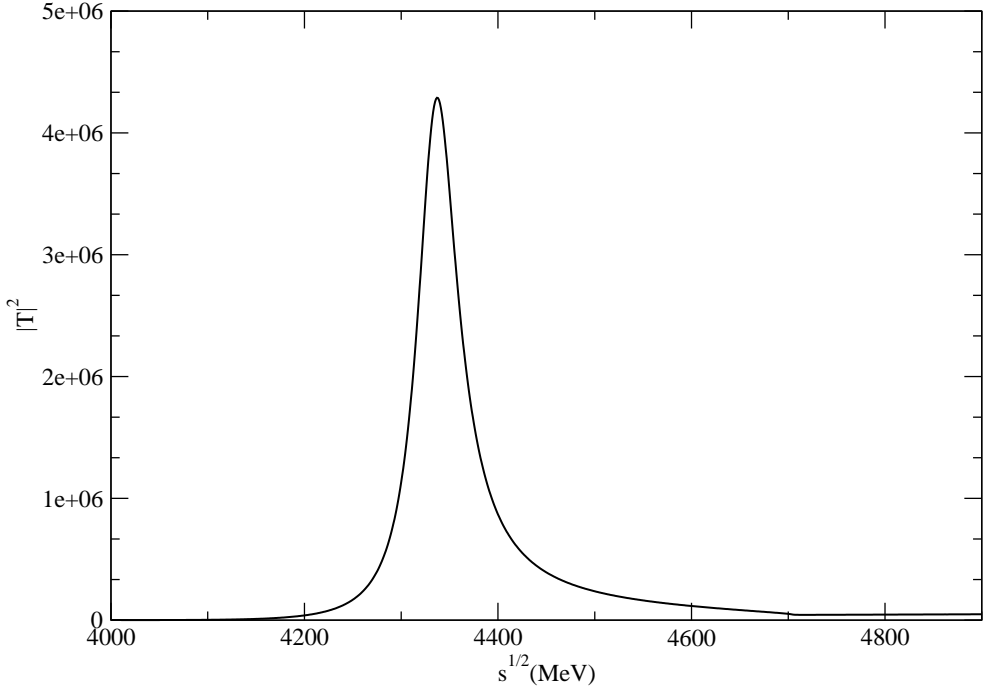}  }
  \caption{Modulus squared of the $ \rho (D^{*}\bar{D}^{*})$ scattering amplitude with total isospin $I=1$.}
  \label{fig:D2strDstar}
  \end{figure}
  \end{center}

In Fig. \ref{fig:D2strDstar} we illustrate the modulus squared $\vert T \vert^{2}$ for the $X \rho \rightarrow X \rho$ scattering as a function of the total energy of the  $ \rho D^{*}\bar{D}^{*}$ system for the case of  $I=1$ and $J=3$. The results show a clear peak at $\sqrt{s}=4338 $ MeV about $360$ MeV below the threshold of the $X(3915)- \rho$ system. The width of the peak is about $50$ MeV.

This looks like a strong binding, but we must keep in mind that the vector-vector interaction in $J=2$ is indeed very strong \cite{Molina:2008jw,Geng:2008gx,DsDBs,rhoDstr}. This is why we are studying these superbound states with spins aligned where the spin of any pair is always $J=2$. The obvious thing is that the $D^{*}\bar{D}^{*}$ state is already bound and since the $\rho D^{*}$ also binds to give the $D_{2}^{*}(2460)$, the system $\rho D^{*}\bar{D}^{*} $  with this configuration will necessarily be bound. This would be the case even if the $\rho$ interacted only with one $D^{*}$. In this case we would have a binding of $m_{D^{*}}+m_{\rho}-m_{D_{2}^{*}}=320$ MeV. The binding that we get is $360$ MeV, which means that we have gained extra $40$ MeV binding by the interaction of the $\rho$ with the $\bar{D}^{*}$. This indicates that there is extra binding from the three body molecular structure. This feature is reminiscent of what happens in Quantum mechanics for the problem of a particle in a well of two attractive $\delta$-functions \cite{Griffiths}. For the symmetric solutions, if the two $\delta$'s are separated, the binding energy of the two $\delta$ potential is the binding of a particle in one $\delta$ well. As the two $\delta$ potentials get closer, the particle starts orbiting the two potential wells and the binding energy grows. This is what we observe here, indicating an extra binding from the orbiting of the $\rho$ around the $D^{*}$ and $\bar{D}^{*}$. 

\begin{center}
  \begin{figure}
  \resizebox{0.8\textwidth}{!}{
  \includegraphics{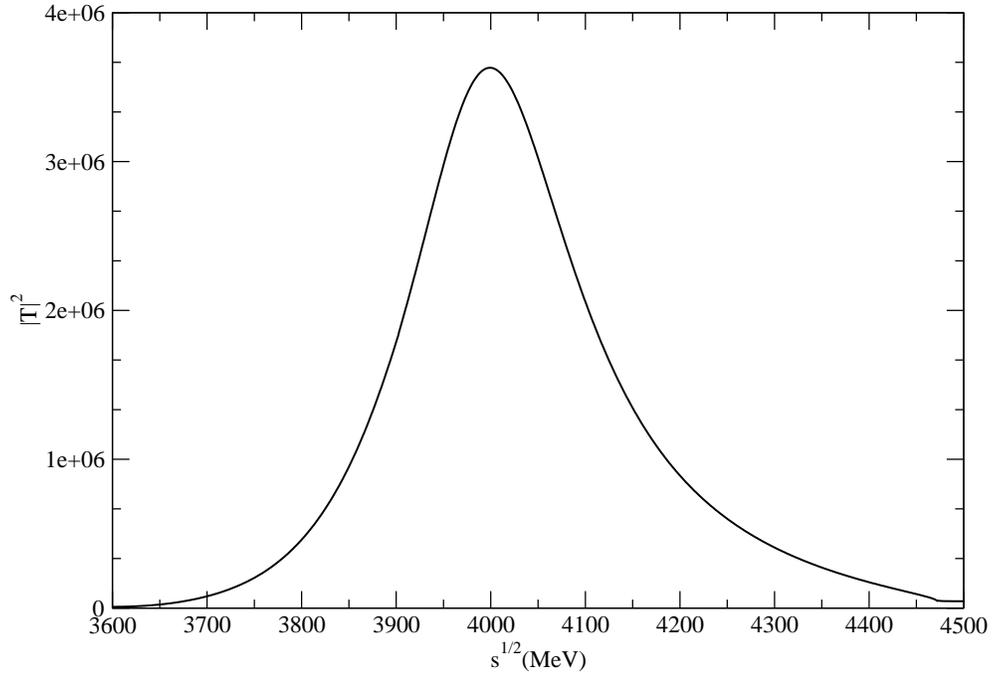}  }
  \caption{Modulus squared of the $ \bar{D}^{*} (\rho D^{*}) $ scattering amplitude with total isospin $I=0$.}
   \vspace{10 mm}
  \label{fig:ampD2sDbsI0}
  \end{figure}
  \end{center}

\begin{center}
  \begin{figure}
  \resizebox{0.8\textwidth}{!}{
  \includegraphics{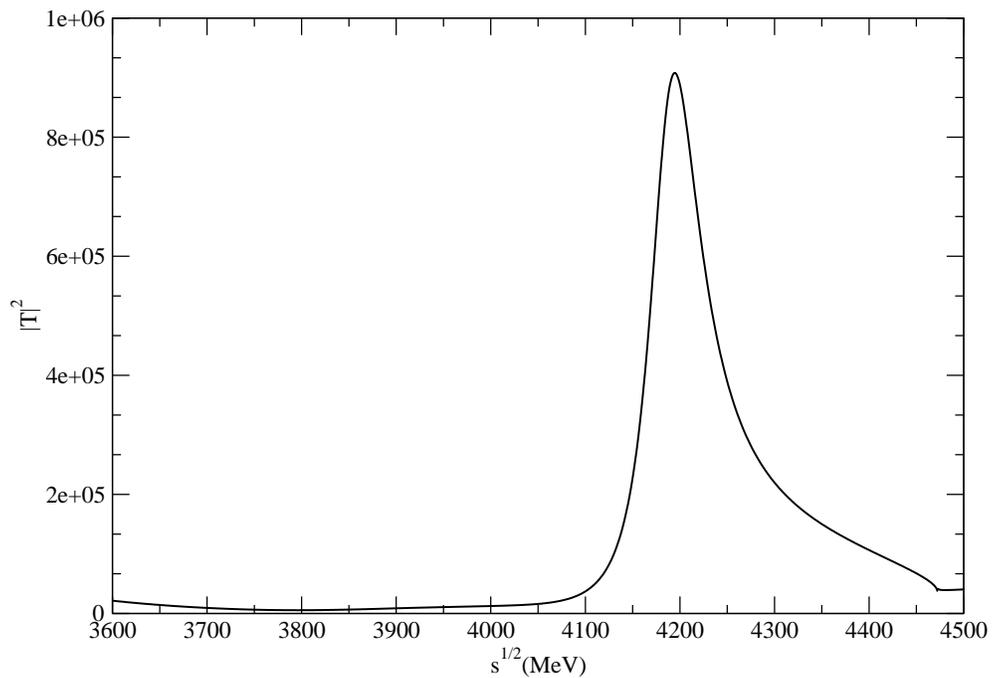}  }
  \caption{Modulus squared of the $\bar{D}^{*} (\rho D^{*}) $ scattering amplitude with total isospin $I=1$.}
  \label{fig:ampD2sDbsI1}
  \end{figure}
  \end{center}

In Figs. \ref{fig:ampD2sDbsI0} and \ref{fig:ampD2sDbsI1} we show the results of $\vert T \vert^{2}$ for the 
case of $D_{2}^{*} \bar{D}^{*} \rightarrow D_{2}^{*} \bar{D}^{*} $ in total isospin $I=0$ and $I=1$, respectively. We find a peak around $4000$ MeV which is about $470$ MeV below the $D_{2}^{*}(2460)$ and $\bar{D}^{*}$ threshold for the  isospin $I=0$ case. The width of this state is quite large about $250$ MeV. If we conduct the same exercise as before, the $D^{*}$ with a $\rho$ would be bound by $320$ MeV, and the $D^{*}$ with the $\bar{D}^{*}$ by about $63$ MeV. We would think that the $D^{*}$ is orbiting the $\rho$ where it is more bound and get some extra binding from orbiting the $D^{*}$. It is not clear why one passes to $470$ MeV binding. It is also unclear why the width is much larger. Probably we have to accept that in this case, since the $D^{*}$ is heavier than the $\rho$, there are limitations to the applications of the FCA, and we should accept this result as an indication that we could now have a state more bound than in the former case, which fulfills all the conditions for a reliable application of the FCA and hence is more reliable, but we cannot be certain about the mass and the width.

 For the isospin $I=1$ case, we see a clear peak around $4195$ MeV,  and the width is around $60$ MeV. 
   The position of the peak is about $270$ MeV below the $D_{2}^{*}(2460)$ and $\bar{D}^{*}$ threshold.  These results are more intuitive than before. We can apply the same argumentation as before, but now according to Eqs. (\ref{eq:Spin1}) and (\ref{eq:Spin2}), we can see that the weight of the $t_{\rho D^{*}}^{I=1/2}$ amplitude in $\langle  \bar{D}^{*} (\rho D^{*}) \vert t \vert \bar{D}^{*}(\rho D^{*}) \rangle $ for $I=0$ is unity while for $I=1$ it is $1/9$, and this is the amplitude that contains the attractions that binds the  $D_{2}^{*}(2460)$. With the caveat about the arguments used before for the case of $I=0$, it looks clear that the binding should be smaller than for the case of $I=0$ and the width is also similar to that of the $\rho(D^{*} \bar{D}^{*} )$ molecule.

 \section{Conclusions}

We studied the interaction of the $\rho$ and $D^*$, $\bar D^*$ with spins aligned using the Fixed Center Approximation to the Faddeev equations. This guarantees that we have $J=2$ for any of the pairs, where the vector-vector interaction is strongest, and leads to three body states with $J=3$.  We first select a cluster of $D^* \bar D^*$, which is found to be bound in $I=0$ and can be associated to the X(3915), and then let the $\rho$ meson orbit around the $D^*$ and $\bar D^*$. In this case the FCA produces an amplitude for $\rho$-$D^* \bar D^*$ scattering which has a clear and narrow peak around 4340 MeV. The case of a $\bar D^*$ orbiting around a cluster of $\rho D^*$ is more uncertain because the mass of the external particle is heavier than the one of the $\rho$ in the cluster, and the FCA is less reliable. In this case the cluster makes the $D^*_2(2460)$ state, and  we point at some qualitative results, with an $I=0$ state around 4000 MeV and an $I=1$ state around 4200 MeV. 
  The results obtained for the $I=1$ state with the $\rho$ orbiting around the X(3915) should be realistic since the $\rho$ is lighter than the constituents of the cluster.  In the other case our results should be taken as indicative, but strong arguments are given that these states should be strongly bound.  

    The results obtained here should serve to stimulate calculations with more accurate three body tools, as those of \cite{MartinezTorres:2007sr,MartinezTorres:2008gy,MartinezTorres:2009xb,MartinezTorres:2008kh,MartinezTorres:2010zv} which could make predictions on these interesting $J=3$ states. One should recall at this point that states with increasing spin number already exist in the light sector \cite{Roca:2010tf} and in the strange sector \cite{YamagataSekihara:2010qk}. What we have done here it to extend this to the hidden charm sector. 
Parallelly, it would also be interesting to investigate states of large spin in the region of mass investigated here. The results obtained in this work provide sufficient support for a devoted search of such states.

\section*{Acknowledgments}  
This work is partly supported by the Spanish Ministerio de Economia
y Competitividad and European FEDER funds under the contract number
FIS2011-28853-C02-01 and FIS2011-28853-C02-02, and the Generalitat
Valenciana in the program Prometeo, II-2014/068. We acknowledge the
support of the European Community-Research Infrastructure
Integrating Activity Study of Strongly Interacting Matter (acronym
HadronPhysics3, Grant Agreement n. 283286) under the Seventh
Framework Programme of EU. This work is also partly supported by TUBITAK under the project No. 113F411.
X.-L.R acknowledges support from the Innovation Foundation of
Beihang University for Ph.D. Graduates and  the National Natural Science Foundation of China under Grant Nos. 11375024.

\end{document}